\documentclass[onecolumn,showpacs,preprintnumbers,amsmath,amssymb,aps,pra]{revtex4}

\usepackage{mathrsfs}
\usepackage{txfonts}
\usepackage{amsmath}
\usepackage{graphicx}
\usepackage{dcolumn}
\usepackage{bm}
\usepackage{ulem}

\begin{document}

\title{Non-Hermitian guided modes and exceptional points using loss-free negative-index materials}

\author{Li-Ting Wu$^{1}$, Xin-Zhe Zhang$^{2}$, Ru-Zhi Luo$^{2}$, and Jing Chen$^{2,3}$} \email{jchen4@nankai.edu.cn}

\address{$1$ School of Information and Communication Engineering, Nanjing Institute of Technology, Nanjing 211167, China\\
$2$ MOE Key Laboratory of Weak-Light Nonlinear Photonics, School of Physics, Nankai University, Tianjin 300071, China \\
$3$ Collaborative Innovation Center of Extreme Optics, Shanxi University, Taiyuan, Shanxi 030006, China
}
\date{\today}

%%%%%%%%%%%%%%%%%%% abstract %%%%%%%%%%%%%%%%

\begin{abstract}
We analyze the guided modes in coupled waveguides made of negative-index materials without gain or loss. We show that it supports non-Hermitian phenomenon on the existence of guided mode versus geometric parameters of the structure. The non-Hermitian effect is different from parity-time ($\mathcal{PT}$) symmetry, and can be explained by a simple coupled-mode theory with an anti-$\mathcal{PT}$ symmetry. The existence of exceptional points and slow-light effect are discussed. This work highlights the potential of loss-free negative-index materials in the study of non-Hermitian optics.
\end{abstract}

\pacs{123 }


\maketitle
%%%%%%%%%%%%%%%%%%%%%%%%%%  body  %%%%%%%%%%%%%%%%%%%%%%%%%%

\section{Introduction}

Recent years the non-Hermitian physics, especially parity-time ($\mathcal{PT}$) symmetry \cite{R01, R02, R03, R04, R05, R06, R07}, has attracted much attention in the society of optics, because it provides an additional degree of freedom in manipulating the dynamics of optical waves. Nevertheless, the requirement on proper strength of gain and loss effects hinders the realistic applications of $\mathcal{PT}$ symmetric optics. The advances of other categories of non-Hermitian optics, e.g. anti-$\mathcal{PT}$ symmetry \cite{R08, R09, R10, R11}, might overcome this drawback because gain and loss are not strictly required.

The simplest configuration in studying $\mathcal{PT}$ symmetric optics is two parallel waveguides (WGs). Such a kind configuration can be readily studied by using Maxwell's equations, and it supports some intriguing optical effects especially the stopped light at exceptional points (EPs) \cite{R12, R13, R14, R15} that separate the conserved and broken $\mathcal{PT}$ phases \cite{R16, R17, R18, R19, R20}. However, for an optical wave propagating inside a WG, it contains multiple important physical parameters especially the wavevector $k$ charactering the propagation of phase front and the Poynting vector $S$ presenting the propagation of energy flux. They carry different informations about the wave, and might have different directions of propagation \cite{R21, R22}. The extreme situation is that in a negative-index material (NIM) with simultaneous negative magnetic permeability ($\mu_\text{NIM}<0$) and electric permittivity ($\epsilon_\text{NIM}<0$), where the direction of $k$ is opposite to that of $S$ \cite{R23, R24, R25, R26, R27}. However, non-Hermitian feature of coupled WGs containing NIM has not been extensively discussed. Only recently do we notice that Mealy etc. al. \cite{R28} have showed that EP of degeneracy can be obtained in two coupled WGs by a proper coupling of forward and backward waves, where the backward waves are generated by a proper designed grating.

In this article, we check the optical waves in coupled WGs made of NIM and positive-index material (PIM). A PIM has positive magnetic permeability and electric permittivity, and covers all the dielectric materials in nature. We show that the guided modes in this system also support non-Hermitian feature, even when all the constituent media are loss-free. At a given angular frequency and WG thicknesses, the wavevector $k$ of the eigenmodes and the associated eigenvectors vary with the distance $a$ between the two WGs, and they coalesce at a critical value of $a_c$ below which the wavevector $k$ becomes complex. This phase transition point is an EP. A simple non-Hermitian coupled-mode theory is utilized to explain our results, which implies that the field dynamics induced by NIM belongs to anti-$\mathcal{PT}$ symmetry. Features at EPs including the slow-light effect are discussed.This study highlights the important novelties of NIM, especially its great potential in the study of non-Hermitian optics by bypassing many restrictions of $\mathcal{PT}$ symmetry.

\section{Simulation and analysis}

Let us consider the simple structure shown in Fig. 1. It contains two straight WGs made of lossless media surrounded by air. One WG is PIM, and we assume that it is a dielectric of $\epsilon_\text{PIM}=4$ and $\mu_\text{PIM}=1$. The other WG is a NIM that has been discussed by various authors \cite{R29, R30, R31}. The well documented structure of NIM with $\epsilon_\text{NIM}=1-\omega_p^2/\omega^2$ and $\mu_\text{NIM}=1-F\omega^2/(\omega^2-\omega_0^2)$ can be utilized here, where $\omega_p=2\pi\times10$ GHz, $\omega_0=2\pi\times4$ GHz, and $F=0.56$ \cite{R29, R30, R31}. An effective NIM is achieved between 4 GHz and 6 GHz.

Because NIM requires an intrinsic dispersion of $\partial (\epsilon_\text{NIM}\omega) / \partial \omega>0$ and $\partial (\mu_\text{NIM}\omega) / \partial \omega>0$ so as to give a positive energy density \cite{R21, R22}, in our below analysis we would keep the angular frequency $\omega$ a constant, and test the variation of the wavevectors $k$ of the eigenmodes versus a geometric parameter of the coupled WGs. Here we set $\omega=2\pi\times5$ GHz  (a free-space wavelength of 6cm), and the dispersion gives $\epsilon_\text{NIM}=-3$ and $\mu_\text{NIM}=-0.556$. The index of refraction equals $n_\text{NIM}=\sqrt{\epsilon_\text{NIM}}\sqrt{\mu_\text{NIM}}=-1.291$. Thickness $b$ of each WG is 4 cm.  The distance $a$ between the two WGs is chosen as the variable.

\begin{figure}
\centerline{\includegraphics[width=10cm]{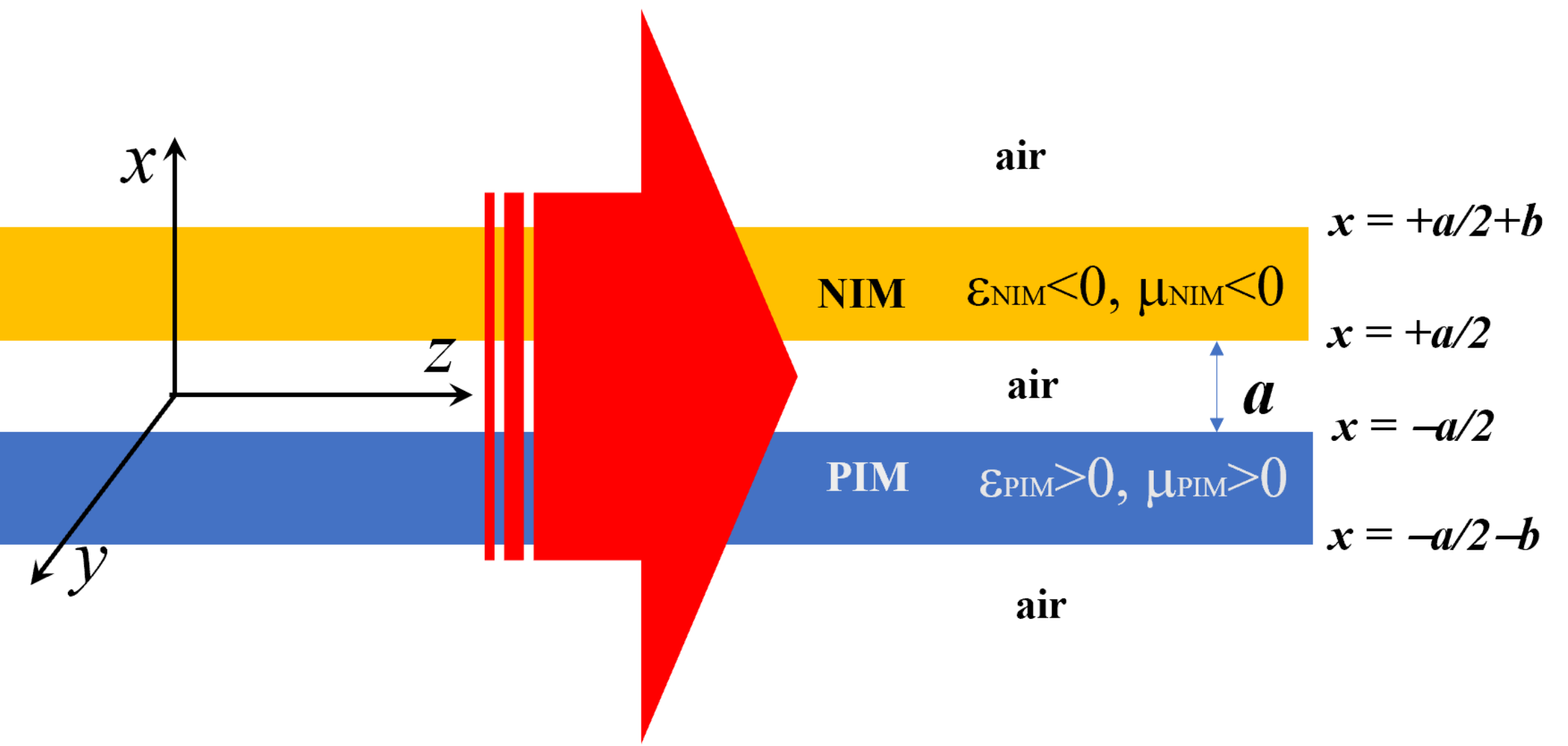}} \caption{Schematic of the configuration under investigation, which contains two straight WGs made of lossless media. One medium is a dielectric with $\epsilon_\text{PIM}>0$ and $\mu_\text{PIM}=1$, while the other one is a NIM with $\epsilon_\text{NIM}=-3$ and $\mu_\text{NIM}=-0.556$.}
\end{figure}

Assuming the direction of the wavevector $k$ is $z$, the dispersion and distribution of the eigenmodes can be numerically found by using Maxwell's equations and boundary conditions \cite{R14, R21, R22}. For example, considering the transverse-electric (TE) polarized eigenmodes with non-vanishing $E_y$ component
\begin{equation}
E_y= e^{-jkz+j\omega t}\left\{
\begin{array}{cc}
  E_1e^{-\beta(x-a/2-b)}, &x>a/2+b\\
  E_2e^{-j\alpha_\text{NIM} (x-a/2)}+E_3e^{+j\alpha_\text{NIM} (x-a/2)}, &a/2+b>x>a/2 \\
  E_4e^{+\beta x}+E_5e^{-\beta x}, &a/2>x>-a/2 \\
  E_6e^{-j\alpha_\text{PIM} (x+a/2)}+E_7e^{+j\alpha_\text{PIM}(x+a/2)}, &-a/2>x>-a/2-b \\
  E_8e^{+\beta(x+a/2+b)}, &x<-a/2-b\\
  \end{array}\right.
\end{equation}
with
\begin{equation}
k^2-\beta^2=\omega^2/c^2,
\end{equation}
\begin{equation}
k^2+\alpha_m^2=\epsilon_m\mu_m\omega^2/c^2,
\end{equation}
where the subscript $m$ stands for NIM and PIM. The associated magnetic fields can be found from Eq. (1) using $\nabla\times \vec{E}=-\partial \vec{B}/\partial t$. By applying the electromagnetic boundary conditions, and defining
\begin{equation}
F_m=\frac{\beta+j\alpha_m/\mu_m}{\beta-j\alpha_m/\mu_m},
\end{equation}
\begin{equation}
\Upsilon_m=\frac{F_m\exp(j2\alpha_m b)-F_m^{-1}}{\exp(j2\alpha_m b)-1},
\end{equation}
it is readily to show that the eigensolutions $k$ are given by
\begin{equation}
\Upsilon_\text{NIM}\Upsilon_\text{PIM}-\exp(-2\beta a)=0.
\end{equation}
By substituting the eigensolutions $k$ back into Eq. (1) we can calculate the distributions of fields. Formula for transverse-magnetic (TM) polarized eigenmodes can be developed similarly.

\begin{figure}
\centerline{\includegraphics[width=13cm]{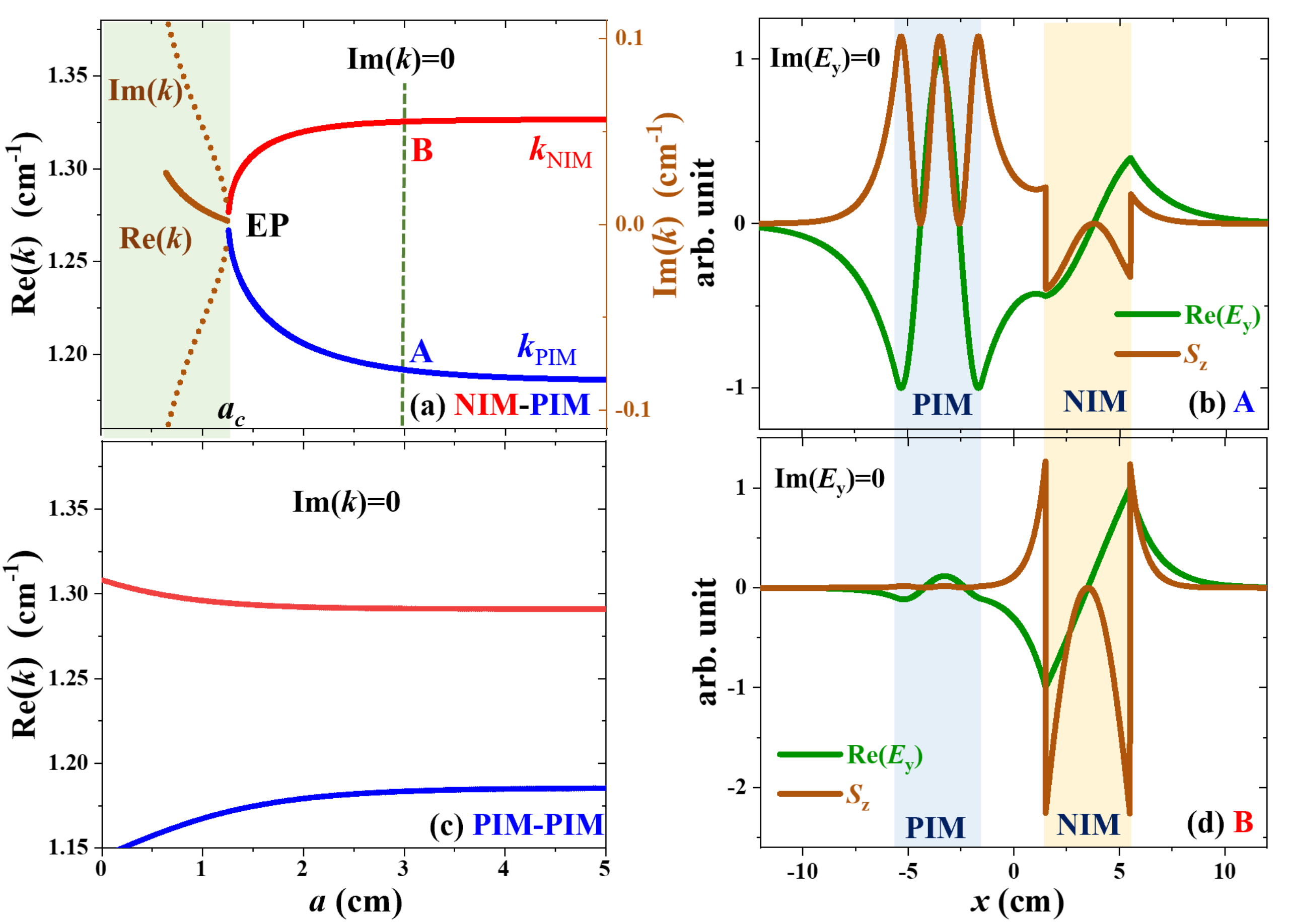}} \caption{(a) Variation of $k$ versus $a$. (b) and (d) are the distributions of field $E_y$ and Poynting vector $S_z$ at the two dispersion points in (a) where $a=3$ cm. (c) shows the results in a PIM-PIM structure. Note that $\text{Im}(E_y)$ is zero in (b) and (d), and the $y$-coordinate of $\text{Im}(k)$ in (a) is shown at the right side, which is properly chosen so that the origins of the curves $\text{Im}(k)$ overlap with EP.}
\end{figure}

Equation (6) can be numerically solved, for example, by calculating the left-hand side of it at different $k$ values to find zeroes. A standard result of $k$ versus $a$ is shown in Fig. 2(a). We can see when the WG distance $a$ is large enough so that the two WGs are weakly coupled, two separated eigenmodes can be achieved. Each eigenmode represents a localized mode in a single WG. From the distributions of field $E_y$ and Poynting vector $S_z$ shown in Figs. 2(b) and 2(d) we can conclude that the eigenmode at $k_\text{NIM}=1.328$ cm$^{-1}$ is supported by the NIM WG, where the direction of energy flux is opposite to that of $k$. The other one at $k_\text{PIM}=1.186$ cm$^{-1}$ is localized at the PIM WG. From the number of nodes ($E_y=0$) inside each WG we can see the field supported by the planar PIM WG is a TE$_2$ mode, and that in the NIM WG is a TE$_1$ one \cite{R30}. Note that the wavevector $k$ is real here, and we have assumed that the expansion coefficient $E_1$ in Eq. (1) is real. As a result, the imaginary component $\text{Im}(E_y)$ is zero and is not shown in Figs. 2(b) and 2(d). The coefficient $\exp(-jkz+j\omega t)$ in Eq. (1) is neglected when plotting the distribution of fields.

Figure 2(a) displays a unique phenomenon that is absent in ordinary dielectric WGs. When $a$ decreases, the modes in the two WGs can couple together to form hybrid eigenmodes. However, here the $k$ values of the eigenmodes approach to each other. When $a$ is smaller than a critical distance $a_c$, the eigenmodes coalesce. Below $a_c$ only complex $k$ exists (brown line and dots in Fig. 2(a)), which can be found by searching for the solutions of Eq. (6) in the space formed by $\text{Re}(k)$ and $\text{Im}(k)$. This phenomenon is in sharp contrast with the ordinary belief about coupled PIM-PIM WGs, that when $a$ decreases, the coupling between the two WGs would become stronger and the split in $k$ should increase rather than decrease. To illustrate this difference, we show an example about $k$ versus $a$ in a PIM-PIM configuration in Fig. 2(c), where NIM in the structure of Fig. 1 is replaced by a conjugate PIM with $\epsilon_\text{PIM}=3$ and $\mu_\text{PIM}=0.556$. We can see in this PIM-PIM configuration all the solutions of $k$ are real, and the two branches of $k$ do not coalesce.

\begin{figure}
\centerline{\includegraphics[width=13cm]{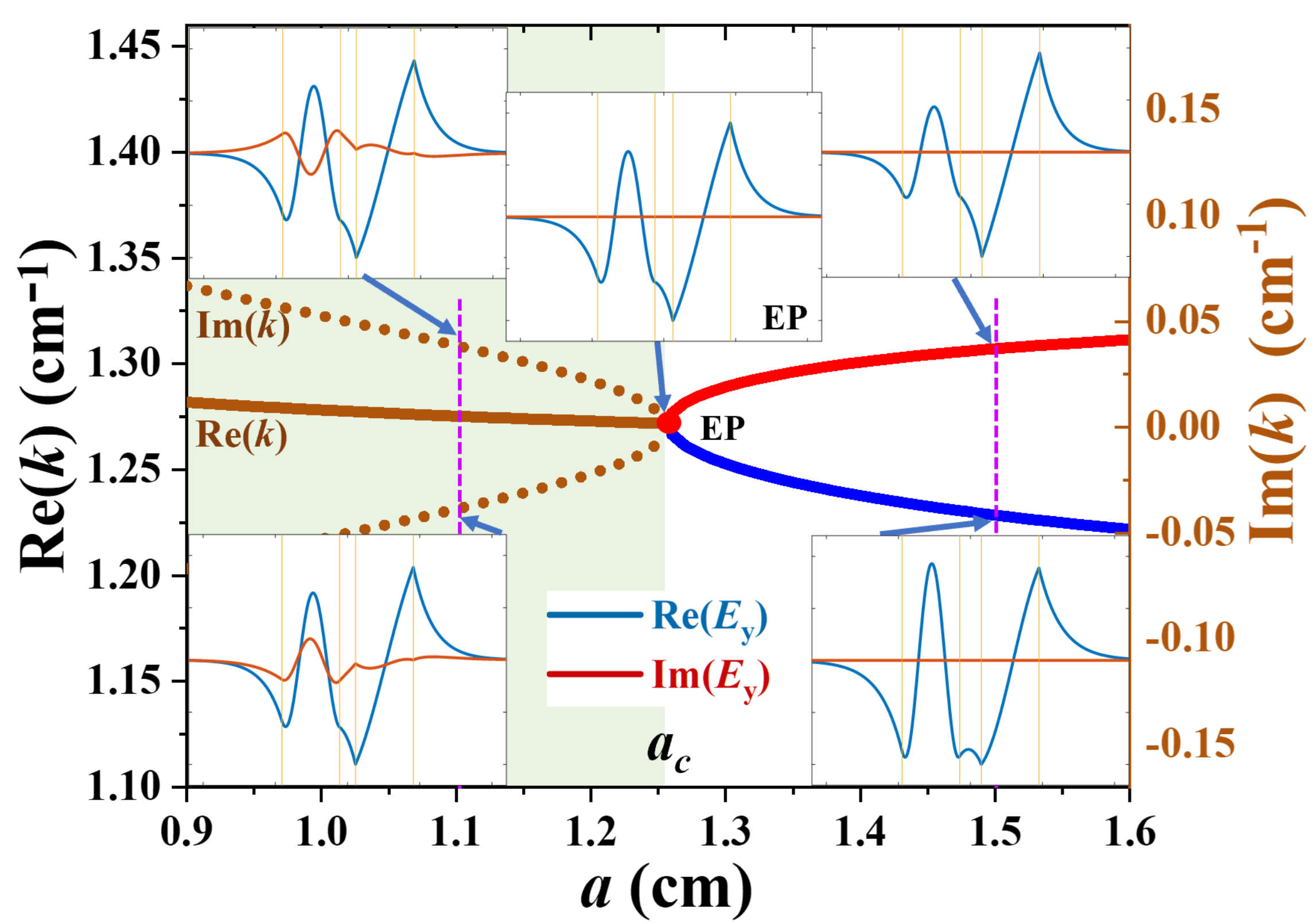}}
\caption{Distributions of fields $E_y$ at some different points in the $k$ versus $a$ curves. They confirm that the transition point at $a_c$ is an EP. Note that $\text{Im}(E_y)$ is no longer zero when $a<a_c$ where $\text{Im}(k)\neq0$.}
\end{figure}

The coalescence of dispersion at a critical value of $a_c$ shown in Fig. 2 resembles EPs in $\mathcal{PT}$ symmetry very much \cite{R12, R13, R14, R15, R16, R17, R18, R19, R20}. We check the eigenmodes around this point (see Fig. 3), and find that when approaching this point from $a>a_c$, the field distributions of the two eigenmodes become more and more similar with each other. So at this critical point not only the eigensolutions $k$ but also the associated eigenvectors coalesce simultaneously. This point is not a diabolic point in Hermitian system, but a standard EP \cite{R16, R17, R18}. The region of $a>a_c$ ($a<a_c$), where the eigensolutions $k$ are real (complex), possesses an exact (broken) non-Hermitian phase.

Figure 3 also displays the distributions of fields when $a<a_c$. In this phase-broken region, at each given $a$ value we can generally find two complex solutions of $k$, which are conjugate to each other. The complex solutions of $k$ render complex values of $\beta$ and $\alpha_m$ given by Eqs. (2) and (3). Consequently,  $\text{Im}(E_y)$ is no longer zero (see the insets of Fig. 3).

To check how the curves of $k$ versus $a$ shown in Fig. 2(a) vary when parameters of the coupled WGs change, we repeat the calculations at different $\epsilon_\text{PIM}$ values. It would change the $k_\text{PIM}$ value of the PIM WG. One can also achieve the same effect by changing the thickness of it. As for the NIM WG, we would keep all the parameters constant so that $k_\text{NIM}$ does not vary in this article.

Figure 4 displays the results when $\epsilon_\text{PIM}$ increases linearly from 4.3 to 4.6. The increased $\epsilon_\text{PIM}$ would push the $k_\text{PIM}$ value in the PIM WG close to or even pass $k_\text{NIM}$ in the NIM WG.  From Fig. 4 we can see with $\epsilon_\text{PIM}$ increasing, the loop of $k$ versus $a$ would shrink, and eventually disappear around $\epsilon_\text{PIM}=4.5$. When $\epsilon_\text{PIM}$ further increases from 4.5, the curve appears again and becomes larger. Since $k_\text{NIM}=1.328$ cm$^{-1}$ does not change, the scenario of $\epsilon_\text{PIM}=4.5$ in fact is around the critical point where the degeneracy of $k_\text{PIM}=k_\text{NIM}$ takes place.

\begin{figure}
\centerline{\includegraphics[width=13cm]{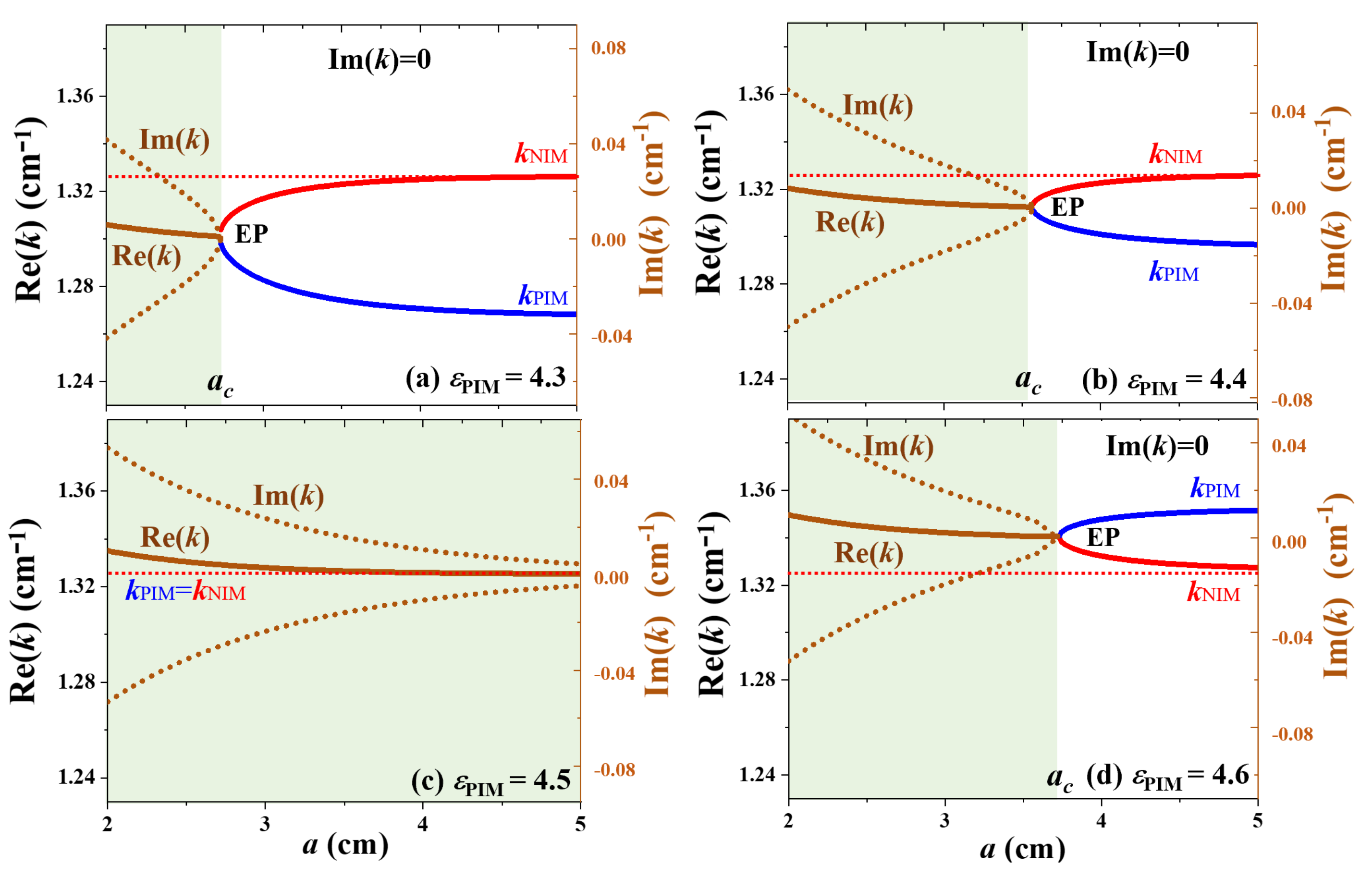}} \caption{Variation of $k$ versus $a$ at different $\epsilon_\text{PIM}$ values, from (a) $\epsilon_\text{PIM}=4.3$ to (d) $\epsilon_\text{PIM}=4.6$.}
\end{figure}

\section{A non-Hermitian coupled-mode theory}

In the above simulation all the media in the coupled WGs are loss-free. Nevertheless, the coalescence of eigenmodes at a critical value of $a_c$ hints that this loss-free system can possess non-Hermitian phenomenon. Here we could utilize a non-Hermitian coupled-mode theory \cite{R21} in the form of $M\Psi=k_\pm\Psi$ to explain the results, which is

\begin{equation}
\left[
\begin{array}{cc}
  k_\text{NIM}  & -\gamma \\
  \gamma & k_\text{PIM} \\
  \end{array}\right]
\left[
\begin{array}{cc}
   \psi_\text{NIM} \\
   \psi_\text{PIM} \\
  \end{array}\right]=
k_{\pm}\left[
\begin{array}{cc}
   \psi_\text{NIM} \\
   \psi_\text{PIM} \\
  \end{array}\right].
\end{equation}
Here $k_\text{NIM}$ and $k_\text{PIM}$ are the wavevectors of guided modes in separated NIM and PIM WGs, respectively, at the given angular frequency $\omega$. Both $k_\text{NIM}$ and $k_\text{PIM}$ are real. Parameter $\gamma$ is also real and characterizes the mutual coupling between the two WGs. The basic vectors $\psi_\text{NIM}$ and $\psi_\text{PIM}$ represent field components of the guided modes in characterizing the eigenvectors. Evidently, Eq. (7) is non-Hermitian because the off-diagonal elements are not conjugates, i.e. $M_{12}\neq M_{21}^*$. The choice of $M_{21}=-M_{12}=\gamma$ can be explained by simultaneously considering two aspects. One is that the total energy should be conserved in this loss-free system \cite{R28, R32}. The other is that the propagating directions of energy in the NIM and PIM WGs are opposite to each other, so the coupling between them is contradirectional \cite{R28, R32}.

To see whether Eq. (7) can explain the main features in our former analysis, we can first solve it. Defining
\begin{equation}
k_\text{NIM} =k_0+\Delta,
\end{equation}
\begin{equation}
k_\text{PIM} =k_0-\Delta,
\end{equation}
where $k_0$ is the mean wavevector, and $\Delta$ is the detuning, the eigen-solutions $k_\pm$ are
\begin{equation}
k_{\pm}=k_0\pm \sqrt {\Delta^2-\gamma^2}.
\end{equation}

By using Eq. (10) we can fit the curves $k$ versus $a$ given by Eq. (6) and find how the magnitude of $\gamma$ varies with the WG distance $a$. Results of $\epsilon_\text{PIM}=4$ is shown in Fig. 5. When performing the best fitting, we firstly choose proper values of $k_\text{NIM}$ and $k_\text{PIM}$, which are $1.328$ cm$^{-1}$ and $1.186$ cm$^{-1}$, respectively, for the results shown in Fig. 5. The values of $k_\text{NIM}$ and $k_\text{PIM}$, as well as those of $k_0=1.257$ cm$^{-1}$ and $\Delta=0.071$ cm$^{-1}$, are kept as constants. Then, based on the $k_{\pm}$ values obtained from Eq. (6) we can use Eq. (10) to find the values of $\gamma$. Because the two solutions of Eq. (6) might give different values of $\gamma$, we would only keep the averaged one, which is then substituted back into Eq. (10) to test the discrepancy from Eq. (6). Similar fitting process is also performed in the phase-broken region of complex $k$ values.

From Fig. 5(a) we can see the prediction from Eq. (10) fits well with those of Eq. (6). The deviation between $\text{Re}(k)$ in the phase-broken region might be fixed by assuming a spatial dependence of $k_\text{NIM}$ and $k_\text{PIM}$ on $a$, which would not be discussed here. It is interesting to emphasize that the magnitude of $\gamma$ exponentially decreases with increasing $a$, and can be approximately expressed by
\begin{equation}
\gamma=\gamma_0\exp(-a/L_c)
\end{equation}
where $\gamma_0=0.18$ cm$^{-1}$. The decay length $L_c$ equals 1.45 cm, very close to the value of 1.438 cm given by $\beta^{-1}=(k_0^2-\omega^2/c^2)^{-1/2}$. Evidently, when $a$ is smaller than $a_c$, the mutual coupling strength $|\gamma|$ is stronger than $|\Delta|$, then the phase of the coupled WGs is spontaneously broken and gives complex $k$ values. When $a$ is larger than $a_c$, $|\gamma|$ becomes smaller than $|\Delta|$, and the phase is conserved. Position $a_c$ is the critical phase transition point of EP where $|\gamma|=|\Delta|$. When the two WGs are far away from each other, $\gamma$ is zero, and $k_{\pm}=k_\text{NIM, PIM}$.  Now the eigenvectors are $\Psi=[1,0]^T$ and $\Psi=[0,1]^T$, and the fields are localized in separated WGs.

\begin{figure}
\centerline{\includegraphics[width=13cm]{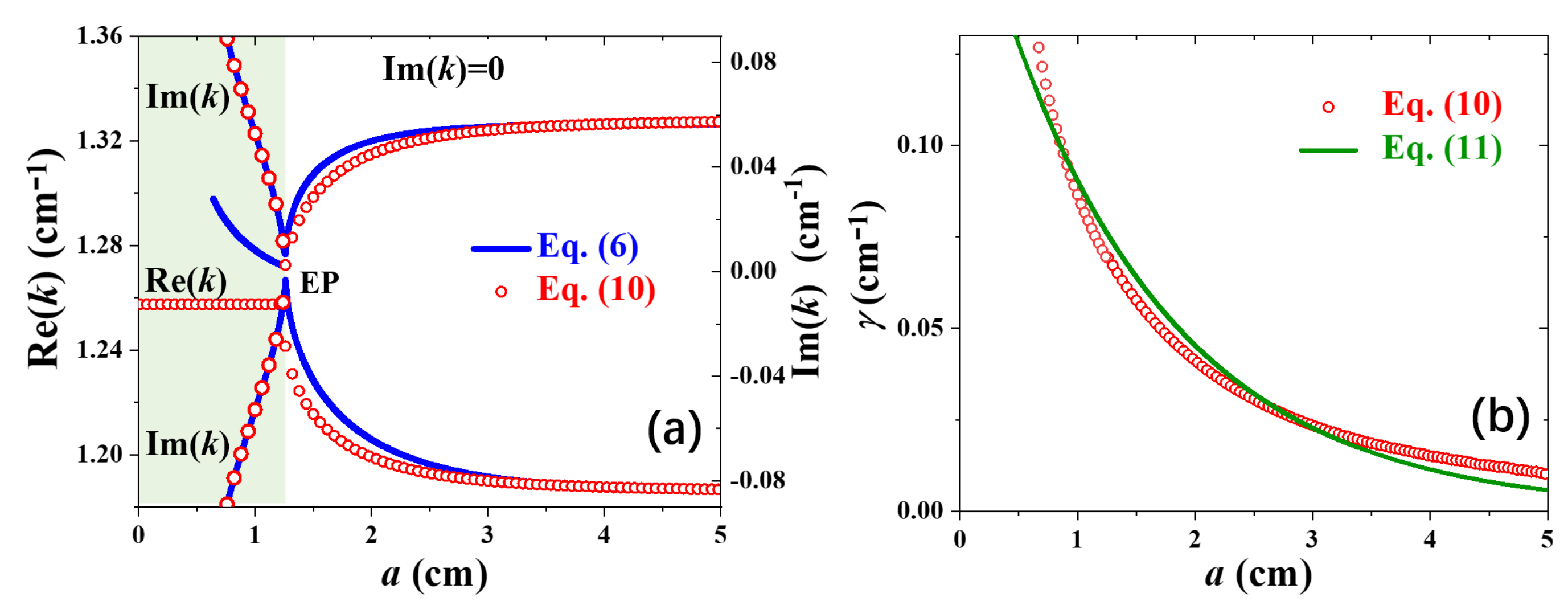}}
\caption{Variations of (a) $k$ and (b) $\gamma$ versus the WG distance $a$ by using Eq. (10) to fit the results from Eq. (6).  Green line in (b) is an exponential fit of $\gamma$ using Eq. (11). }
\end{figure}

Furthermore, the detuning $\Delta$ determines the size of the loop in the $k$ versus $a$ spectrum. When $k_\text{NIM}=k_\text{PIM}$, i.e. the eigensolutions in the NIM and PIM WGs are degenerated, the phase is always spontaneously broken because now $\Delta=0$ and any nonzero $\gamma$ would give complex $k_\pm$ values. It is the case shown in Fig. 4(c) at $\epsilon_\text{PIM}=4.5$. Away from this degeneracy scenario, $|\Delta|$ increases and the loop becomes more and more great, which can be observed in other plots in Fig. 4.

Now, we can pay attention to the EP where
\begin{equation}
\Delta=\pm\gamma
\end{equation}
that gives $k_{\pm}=k_0$.
The eigenfuntion of the eigenmode is
\begin{equation}
\left[
\begin{array}{cc}
   \psi_\text{NIM} \\
   \psi_\text{PIM} \\
  \end{array}\right]_\text{EP}=
\left[
\begin{array}{cc}
1 \\
\text{sign}(\Delta/\gamma) \\
  \end{array}\right].
\end{equation}
where the function $\text{sign($x$)}$ equals 1 ($-1$) when  $x>0$ $(x<0)$. In other words, at this EP the fields in the two WGs are in-phase (a phase difference of zero) or out-phase (a phase difference of $\pi$) with each other. The exact phase difference is determined by the sign of $\Delta/\gamma$. Generally $\gamma$ is determined by the overlap integral of fields in the gap between the two WGs and its sign is fixed \cite{R14, R15, R28},  but $\text{sign}(\Delta/\gamma)$ can be tuned by changing the resonant conditions of $k_\text{NIM, PIM}$ in the two WGs, e.g. by changing the index of refraction inside or the thicknesses of WGs. As a consequence, the eigenvetcor at EP is tunable.

Note that here we could not make any comment on the amplitudes of fields in the two WGs because the refractive indexes in them are not required to be equal. It is in sharp contrast with $\mathcal{PT}$ symmetric WGs that generally only symmetric configurations are considered \cite{R10, R11, R12}. This drawback hinders us to reveal more unique features about EP. However, the phase-jump, as a signature of $\text{sign}(\Delta/\gamma)$ at EP, is an observable, e.g. by choosing two sets of $\epsilon_\text{PIM}$ values so that in one case $\Delta>0$ while in the other case $\Delta<0$, and then analyzing the field patterns inside the structure. The results about the distributions of $E_y$ at EPs when $\epsilon_\text{PIM}=4.4 (\Delta>0)$ and $\epsilon_\text{PIM}=4.6 (\Delta<0)$ are shown in Fig. 6. From the two plots we can see the fields in the NIM WG are almost identical, but the fields in the PIM WG are flipped with respect to each other. It is then evident that the phase between the two WGs are shifted by $\pi$ in the two scenarios. Furthermore, associated with the $\pi$-phase jump there exists a node inside the gap between the two WGs (red arrow in Fig. 6(b)), which implies that in this scenario $\psi_\text{NIM}=-\psi_\text{PIM}$. From the fact of $\Delta<0$ in Fig. 6(b) we can also conclude that $\gamma>0$.

\begin{figure}
\centerline{\includegraphics[width=12cm]{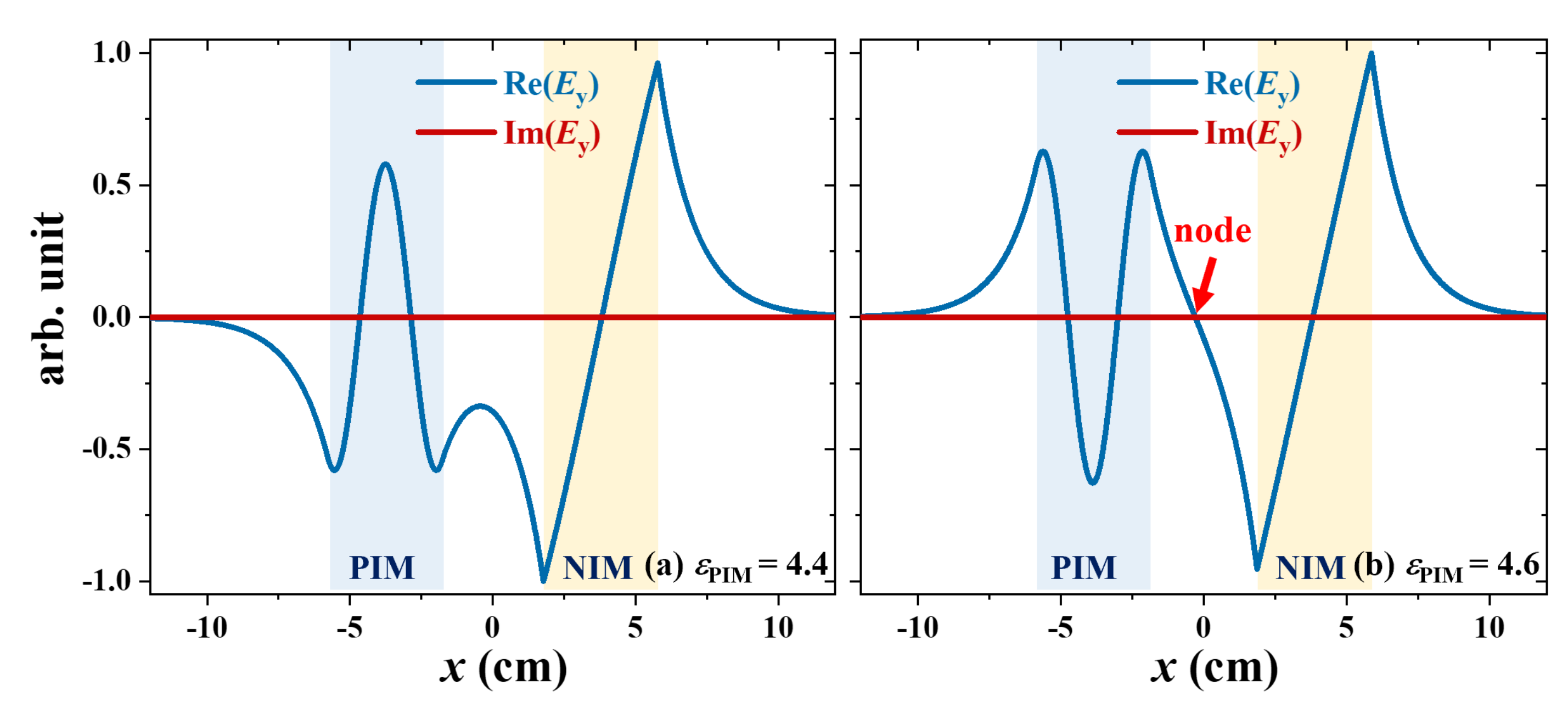}}
\caption{Distributions of fields $E_y$ at EPs when (a) $\epsilon_\text{PIM}=4.4 (\Delta>0)$, and (b) $\epsilon_\text{PIM}=4.6 (\Delta<0)$, respectively.  Red arrow represents the node in (b), which is associated with the $\pi$-phase jump at EP when $\Delta$ varies its sign. }
\end{figure}

The group velocities $v_g$ of the guided modes can also be analyzed. However, since here the angular frequency $\omega$ is kept constant, $v_g$ cannot be found from $\partial \omega/\partial k$.  Here we calculate $v_g$ by using the Poynting vector $S_z$ and the energy density $W$ via $v_g=S_z/W$. When calculating the energy density $W$ we have adapted the formula of  $\epsilon_0\partial (\epsilon_\text{NIM}\omega) / \partial \omega |E|^2+\mu_0\partial (\mu_\text{NIM}\omega) / \partial \omega |H|^2$ \cite{R21, R22}. At 5 GHz, the utilized dispersion gives $\partial (\epsilon_\text{NIM}\omega) / \partial \omega=5$ and $\partial (\mu_\text{NIM}\omega) / \partial \omega=4.975$.

\begin{figure}
\centerline{\includegraphics[width=9cm]{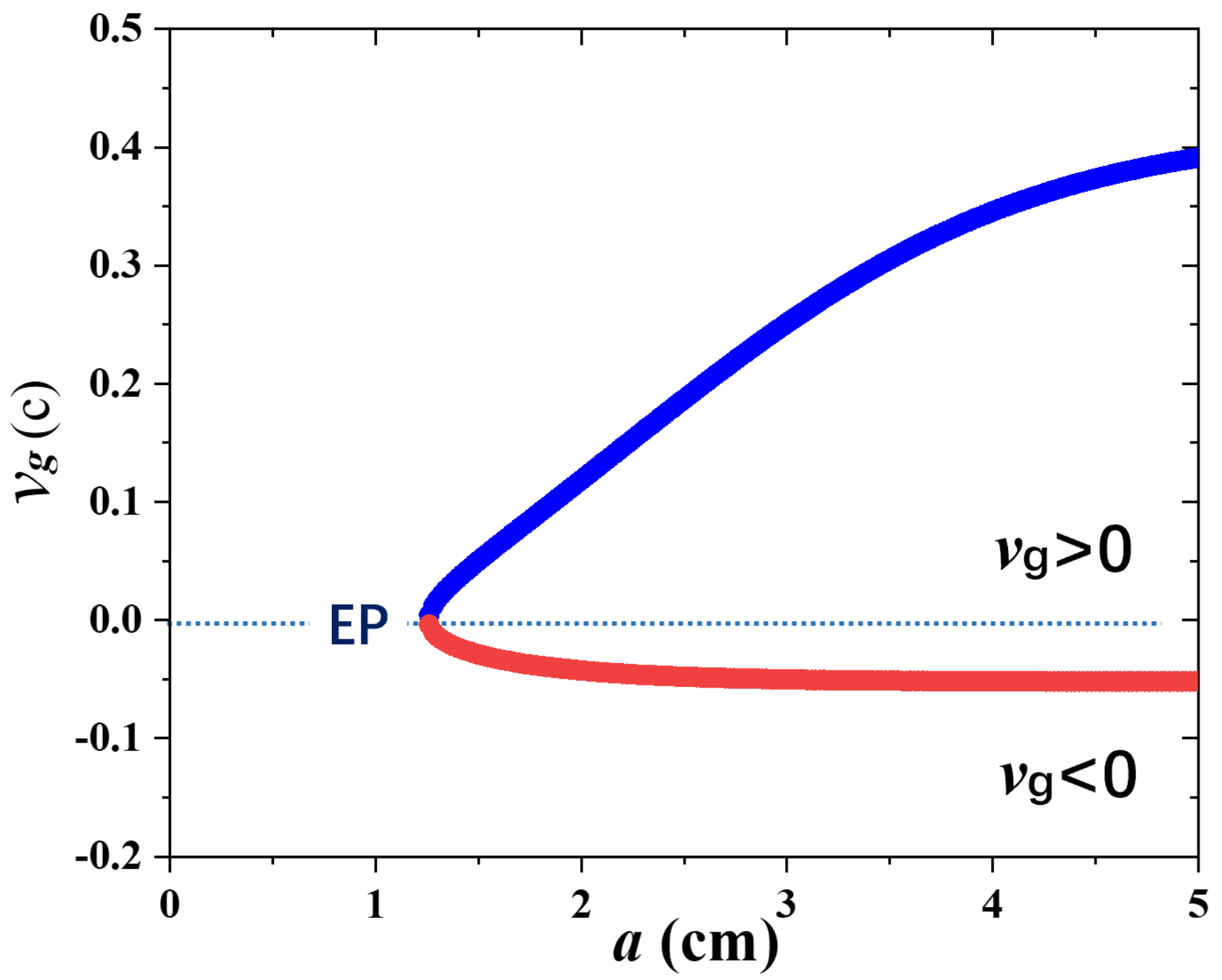}} \caption{Group velocity $v_g$ versus $a$ when $\epsilon_\text{PIM}=4.0$. At EP $v_g$ is zero.}
\end{figure}

The curves of $v_g$ versus $a$ at  $\epsilon_\text{PIM}=4.0$ are shown in Fig. 7. As expected, the values of $v_g$ are limited in the region defined by $v_g=-0.2c$ of NIM (it is negative because the energy flux is backward propagating) and $v_g=0.5c$ of PIM, where $c$ is the speed of light. When the field is mainly localized in the NIM WG, $v_g$ is negative. Otherwise $v_g$ is positive. As $a$ decreases, the split between $k$ also decreases, and $v_g$ of the two branches approach to each other. At EP the group velocity is zero, which can be explained by the balanced positive energy flux in PIM (including the surrounded air) and negative energy flux in NIM, respectively.

Note that the stopped light at EP demonstrated here is different from that in $\mathcal{PT}$ symmetric WGs \cite{R12, R13, R14}. Here the stopped light is associated with the negative flux in NIM, and the whole structure is still Hermitian because no loss or gain is presented. On the contrary, the stopped light in $\mathcal{PT}$-symmetric WGs with spatial distributions of gain and loss cannot be explained by the propagating of energy flux associated with real-valued fields \cite{R14}, but is the consequence of zero $c$-product \cite{R12, R33} of the non-Hermitian system. The $c$-product can be applied when the two WGs are not only geometric but also electromagnetic (except for the loss/gain effect) identical so that $\psi$ in the eigenvector can be unambiguously defined \cite{R12, R33}. Here the two WGs are made of different media, possess different indexes of refraction, and can have different thicknesses, so the applicability of $c$-product needs further discussion.

\section{Discussion}

Above we have shown that non-Hermitian optical effects can be observed in loss-free coupled WGs made of NIMs. Here we would like to make further comments on the category of the non-Hermitian effect. From Eq. (7) we can see it can be classified into the anti-$\mathcal{PT}$ symmetric one because the diagonal elements of the $2\times 2$ matrix $M$ satisfy $\text{Im}(M_{11})=\text{Im}(M_{22})$ and $M_{12}=-M_{21}^*$. It should be emphasized that when utilizing Eq. (7) we have assumed that the basic vectors $\psi_\text{NIM}$ and $\psi_\text{PIM}$ represent field components of the guided modes that characterize the wave functions. The set of analogues of the basic vectors $\psi_\text{NIM} = E_y^\text{NIM}$ and $\psi_\text{PIM}= E_y^\text{PIM}$ is adopted in our analysis. However, since the two WGs are not identical with each other, the choice of the basis vectors is arbitrary. For example, we can also utilize another set of analogues by adding an additional phase of $\pi/2$ to the basic vector $\psi_\text{PIM}$. Now, Eq. (7) becomes
\begin{equation}
\left[
\begin{array}{cc}
  k_\text{NIM}  & j\gamma \\
  j\gamma & k_\text{PIM} \\
  \end{array}\right]
\left[
\begin{array}{cc}
   \psi_\text{NIM} \\
   j\psi_\text{PIM} \\
  \end{array}\right]=
k_{\pm}\left[
\begin{array}{cc}
   \psi_\text{NIM} \\
   j\psi_\text{PIM} \\
  \end{array}\right].
\end{equation}
Equation (14)  is just the standard anti-$\mathcal{PT}$ symmetric operator that has been discussed in many literatures \cite{R08, R09, R10, R11}. It can also explain all the features on the dispersion and distributions of fields. As for which one is preferred in studying NIM, we still prefer Eq. (7) because the node shown in Fig. 6(b) can be more intuitively explained.

The importance of this article is that it proves we can access non-Hermitian optics without using gain or loss. It also shows that the negative energy flux in NIM has many realistic impacts on future applications, especially by considering the fact that all the media are loss-free. Experimental investigation of the non-Hermitian effect discussed here can utilize documented NIM design \cite{R26, R27}, and transfer to other frequency regimes by simply rescaling the geometric parameters of the coupled-WG configuration. Theoretical interest can be paid to discuss the deep-lying physics of the spontaneous phase broken, and the utilization of the anti-$\mathcal{PT}$ symmetry in loss-free NIM to achieve higher-order EPs and mode switching purpose.

Before ending this article, we would like to emphasize again that the whole structure studied here is loss- and gain-free. Consequently, in principle it is a closed Hermitian system, and the energy should be conserved. The non-Hermitian effects discussed above, especially the existence of complex $k$ below $a_c$, should not violate the principle of energy conservation. These complex-$k$ eigenmodes might be associated with some effects similar to the photonic bandgap effect and waveguiding effect below cutoff, which forbid the propagation of optical field without absorbing any energy \cite{R21, R28}. Detailed discussion deserves our further efforts.

\section{Conclusion}

In summary, in this article we check the optical waves in coupled WGs made of lossless NIM and PIM. We show that the guided modes in this kind of gain/loss-free optical system can also support non-Hermitian features. A simple non-Hermitian coupled-mode theory is utilized to explain our results. This theory proves that the critical degeneracy point at $a_c$ is an EP, and the field dynamics induced by NIM belongs to anti-$\mathcal{PT}$ symmetry. Features at EPs including the slow-light effect are discussed. This study highlights the non-negligible novelties of NIM, which can be utilized in studying non-Hermitian optics to bypass many restrictions of $\mathcal{PT}$ symmetry.

\section*{Funding}
Natural National Science Foundation of China (NSFC) (12104227, 12274241); Scientific Research Foundation of Nanjing Institute of Technology (YKJ202021).

\section*{Disclosures}
The authors declare no conflicts of interest.

\section*{Data availability}
Data underlying the results presented in this paper are not publicly available at this time but may be obtained from the authors upon reasonable request.

%%%%%%%%%%%%%%%%%%%%%%% References %%%%%%%%%%%%%%%%%%%%%%%%%

\end{document}